# ON THE TEMPORAL DISTRIBUTION OF CASUALTIES AND DETERMINATION OF MEDICAL LOGISTICAL REQUIREMENTS

M K Lauren

January 2003

# ON THE TEMPORAL DISTRIBUTION OF CASUALTIES AND DETERMINATION OF MEDICAL LOGISTICAL REQUIREMENTS

M K Lauren

Abstract:

It is demonstrated that World War II casualty data display statistical structure that would be expected from multifractal data. Given that the data displayed these properties, it is shown how the existence of power-law tails in the exceedence probability distributions can be used to estimate the likelihood of various casualty levels. Estimates made using this method matched the historical data well.

# EXECUTIVE SUMMARY

**Background**

This study uses ideas developed in previous reports on the theory of using fractals to model combat casualties. This body of work has received significant interest from overseas experts, particularly in the US, UK and Australia. There are now many examples of this work being cited by our collaboration partners, with one notable example being a new book on complexity science and network-centric warfare by a leading UK theorist. This report compliments an earlier report (DTA Report 186) which provided a conceptual framework for understanding these ideas. This paper extends on this work by describing how the properties of fractal data can be used to infer the distribution of casualties in time, and the likelihood of incurring a certain level of casualties within a certain length of time. It is envisaged that these ideas could be incorporated into the models used to determine medical logistics requirements for land operations.

**Sponsor**

CAD Branch, New Zealand Army, under DTA Project 9703.

**Aim**

To obtain equations for estimating the likelihood of given casualty levels during a land operation.

**Results**

It is shown that World War II casualty data display statistical structure and probability distributions that would be expected from fractal data. Given that the data displayed these properties, it was shown how the existence of power-law tails in the exceedence probability distributions could be exploited to allow estimates of the likelihood of various casualty levels. If it can be assumed that all such divisions fight in similar ways, and so obey similar power laws, then one only needs an estimate of the mean daily casualty rate to be able to calculate the probability of a given level of casualties for a particular period. While it appears that the assumption that the power exponent describing the probability distribution does not change (or changes only slightly) from division to division for the data analysed here, it is not clear if this is generally valid. Further investigation is needed on this point.

**CONTENTS**

# 1 INTRODUCTION

The Defence Technology Agency is in the process of developing an Operational Analysis (OA) capability, in keeping with international trends towards using analysis to support acquisition arguments, determine best use of equipment, and to support actual operations. This report deals with the last aspect – support of actual operations.

This short report deals with estimating medical logistics requirements for land warfare. The first part will show how the distribution of casualties in time for a particular operation can be understood using fractal methods, making use of the results of earlier work by the author. The second part will show how the implications of the data being fractal can be used to estimate the probability of different levels of casualties for different periods.

Previous reports hypothesised that casualty distributions are characterised by fractal power laws, so that the statistical structure function is described by an equation of the form:

$$\left\langle \left| B(t_0 + \Delta t) - B(t_0) \right|^2 \right\rangle \propto \Delta t^D \qquad (1)$$

where $B$ is the number of Blue casualties at time $t$, and where $D$ is called the "fractal dimension", which is associated with the distribution of forces on the battlefield (Lauren 1999, 2002a, 2002b). Furthermore, it was hypothesised that the behaviour of the statistical moments of the data were described by an equation of the form:

$$\left\langle \left| \frac{\Delta B}{\Delta t} \right|^q \right\rangle \propto \Delta t^{K(q)} \qquad (2)$$

where $K$ is a non-linear function of the order of the statistical moment, $q$ (Lauren, 2001). The values of $K$ for various $q$ are also called fractal dimensions. Hence the data are described by multiple fractal dimensions, and so are called "multifractal". The validity of these equations was established in an examination of World War II data in Lauren and Stephen (2002). For this report, it is not necessary to understand the properties of fractals in any further depth than to simply understand that fractal data are data that display power laws (see Mandelbrot (1983) for an essay on fractals).

In addition to this, if casualty data do obey such laws, then one would expect to find two things:

Clustered volatility: That is, the most severe casualty days occur around the same times.

Fat-tailed probability distributions: That is, a higher proportion of extremely high casualty days than might be expected from normal or lognormal probability distributions.

The reasons why casualty data should obey these fractal laws are discussed in other papers by the author (Lauren 1999, 2000, 2001, 2002a, 2002b, 2002c), and further investigated by Moffat (Moffat, 2002; Moffat and Passman, 2002; Moffat and Whitty, 2002). Furthermore, it has been noted by other workers that such power laws exist that describe the distribution of casualties for various wars throughout history (Richardson 1941, 1960; Roberts and Turcotte, 1998).

This work does not discuss the underlying mechanisms behind these laws. Indeed, the exact nature of these laws is still unclear, and is only understood in terms of phenomenological arguments. Instead, the purpose here is to show how one might practically apply these results. To do this, the same World War II data as mentioned above are analysed to show how these properties might be used to estimate medical logistics requirements.

## 2. ANALYSIS

Figure 1 shows casualty data (KIA+WIA) for the US 2$^{nd}$ Infantry and Armoured Divisions for the period post the Normandy landings, World War II. It is immediately obvious from visual inspection that the distribution of casualties in time is clustered, as asserted by our hypothesis. This can be quantified by the use of fractal analysis.

It should be expected from Equation 1 that the spectrum of the WWII data obeys a power law, since it can be shown from the Wiener-Khinchine relation that for a power exponent $\beta$, such that $1 < \beta < 3$,

$$|f|^{-\beta} \leftrightarrow |\Delta t|^{\beta-1} \qquad (3)$$

are Fourier transform pairs, where $f$ is frequency.

For the WWII data presented here, this can be seen in Figure 2. Here, the spectra obey approximate straight-line power law slopes, with $\beta$ of just under 1.0 and just over 1.0 for the infantry and armoured divisions respectively.

Though the existence of this power law is evidence that the data is fractal, a more complete characterization can be obtained by examining the multifractal properties of the data (Davis *et al.* (1994) give a particularly good account of this method in the context of analysis of meteorological data).

A multifractal data set has statistical moments that depend on the resolution at which the data is examined, i.e.:

$$\left\langle |C_i|^q \right\rangle \propto \left(\frac{t}{T}\right)^{-K(q)} \qquad (4)$$

where $C_i = \Delta B / \Delta t$ is the number of Blue casualties in the *i*th time step, the angled brackets represent an ensemble average, $t$ is the temporal resolution at which the distribution is being examined, $T$ is the "outer" scale of the scaling range, and $K(q)$ is a non-linear function of the order of the statistical moments, $q$.

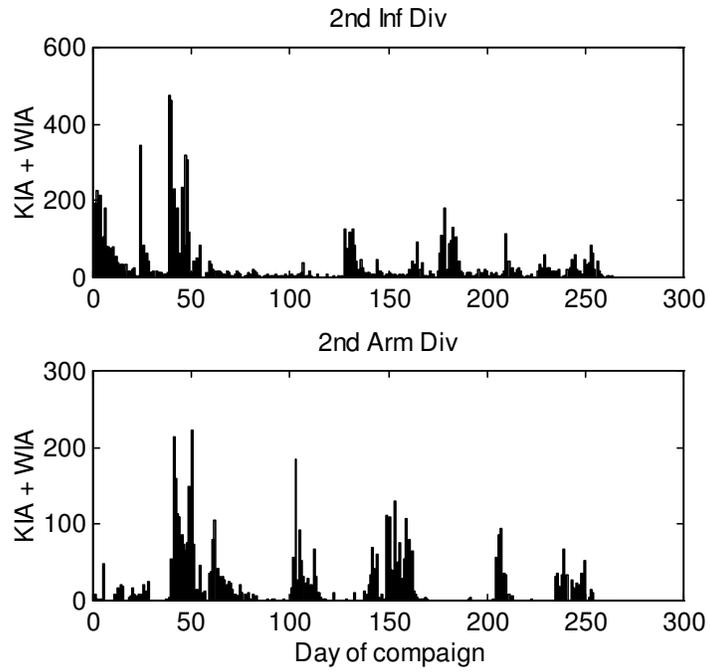

**Figure 1: Daily killed plus wounded figures for the 2$^{nd}$ Infantry and 2$^{nd}$ Armored Divisions.**

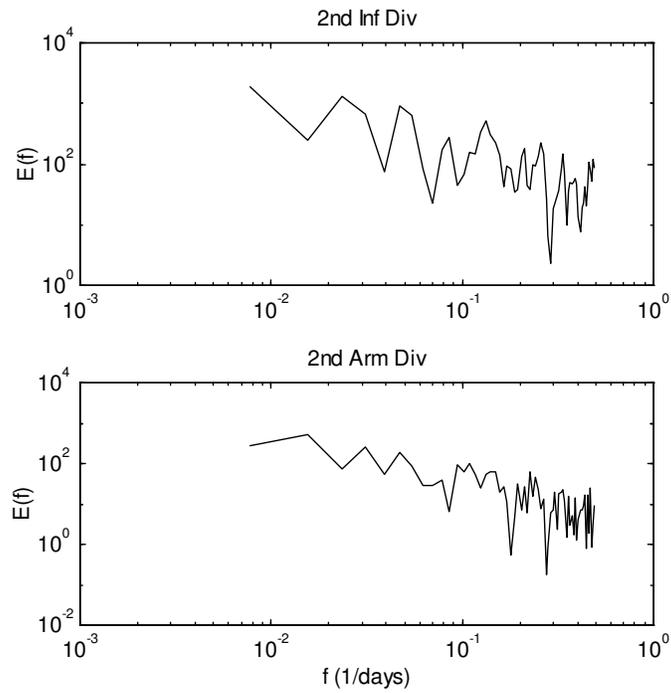

**Figure 2: The Fourier power spectra for the 2$^{nd}$ Infantry and 2$^{nd}$ Armored Division casualty data.**

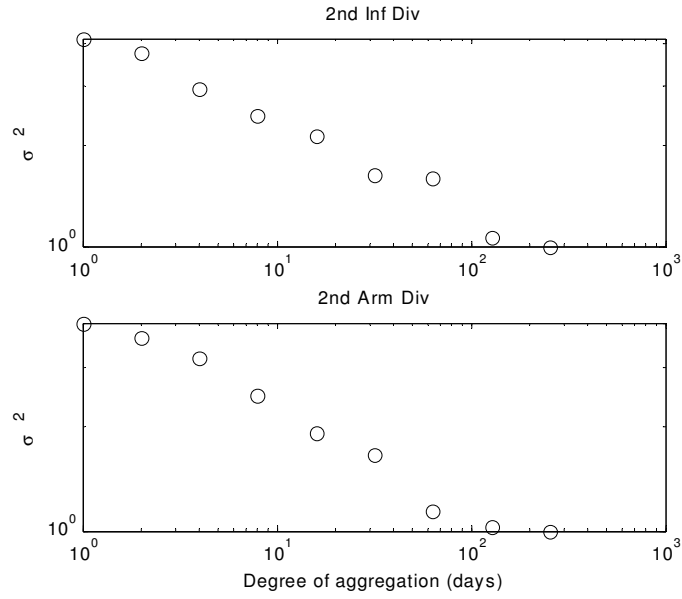

**Figure 3: Dependence of the standard deviation for the casualty data on the degree of data aggregation.**

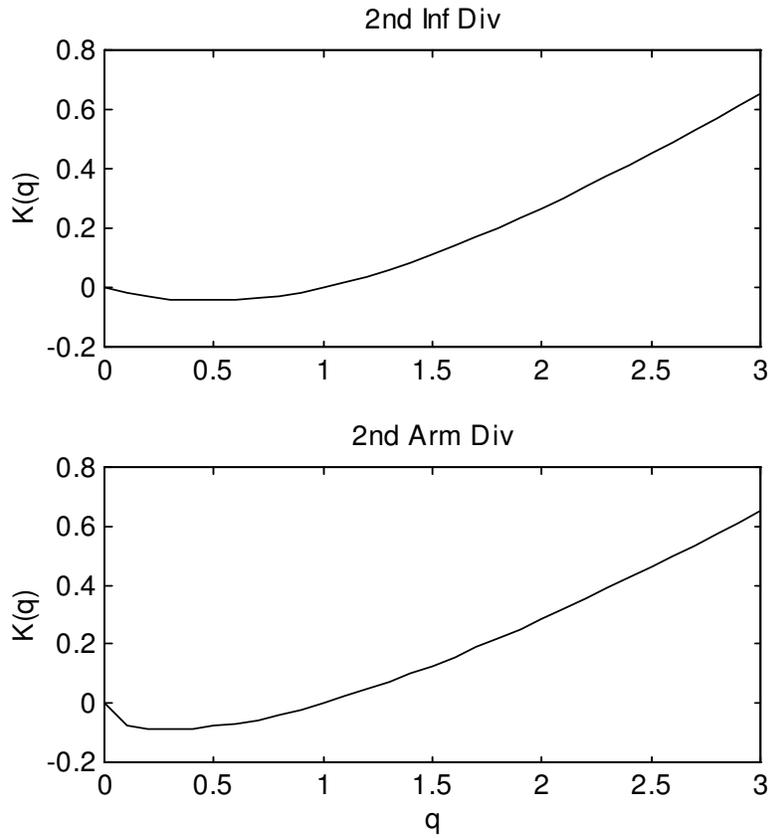

**Figure 4: Scaling exponent *K* as a function of the order *q* of the statistical moments of the casualty data.**

Data are analysed for multifractal behaviour by aggregating neighbouring data points, hence changing $t$. The value of the second-order moment (i.e. $q = 2$) for each set of casualty data can be seen to change with the degree of aggregation in Figure 3. The dependence is a power law, and finding the slope of this gives the value for $K(2)$. The entire $K(q)$ functions for each case can be seen in Figure 4 to be non-linear functions of $q$.

This behaviour for WWII data has been demonstrated in an earlier paper (Lauren and Stephen 2002). In this report, the analysis is taken a step further, to look at the behaviour of the exceedence probability distribution.

**3. ESTIMATING LIKELIHOOD OF EXTREME EVENTS**

Figures 2, 3 and 4 demonstrate that the WWII data discussed here behave like a multifractal, strongly supporting the hypothesis presented in Lauren (1999, 2001, 2002a, 2002b).

This section focuses on medical support logistical estimates. Suppose we want to estimate the likelihood of a given level of casualties for a particular period of time, given a known mean level of casualties for the entire operation. We might do this by using a fractal cascade model to simulate an arbitrarily large number of data points, and obtain the probability distribution from these. It is easy to produce artificial multifractal data that are virtually indistinguishable from real data by using a fractal cascade model. With the right kind of model, it is possible to produce data which give the same $K(q)$ functions as the real data. For such a model, the mean value of the time series data is determined at the start of the cascade, and the other statistical measures of the data are determined by the generator used for the cascade process. Thus such data can be normalized easily to a mean of one, since multiplying the data by a constant does not change the $K(q)$ functions.

A number of simple cascade generators have been suggested for multifractal data. One example is that of Menabde (1997), which uses two parameters. Menabde also describes a method for obtaining the parameters for his model. These parameters are directly related to the fractal dimensions of the data.

The problem of estimating casualty data can be significantly simplified if one makes the assumption that the fractal dimensions of casualty data are the same for all battles. There have been many interesting essays that point out the ubiquitous nature of power laws in complex systems (e.g. Buchanan, 2000). If one were to apply this idea in its broadest philosophical sense to combat data, it would lead to the conclusion that the details of a particular battle are irrelevant to the power law that describes the data, since it is a universal property of warfare. However, in practice it would be unreasonable to claim that any given battle is equivalent to any other. Furthermore, in Lauren (2002c) it is shown how collections of different kinds of battles lead to different estimates of the power law exponent.

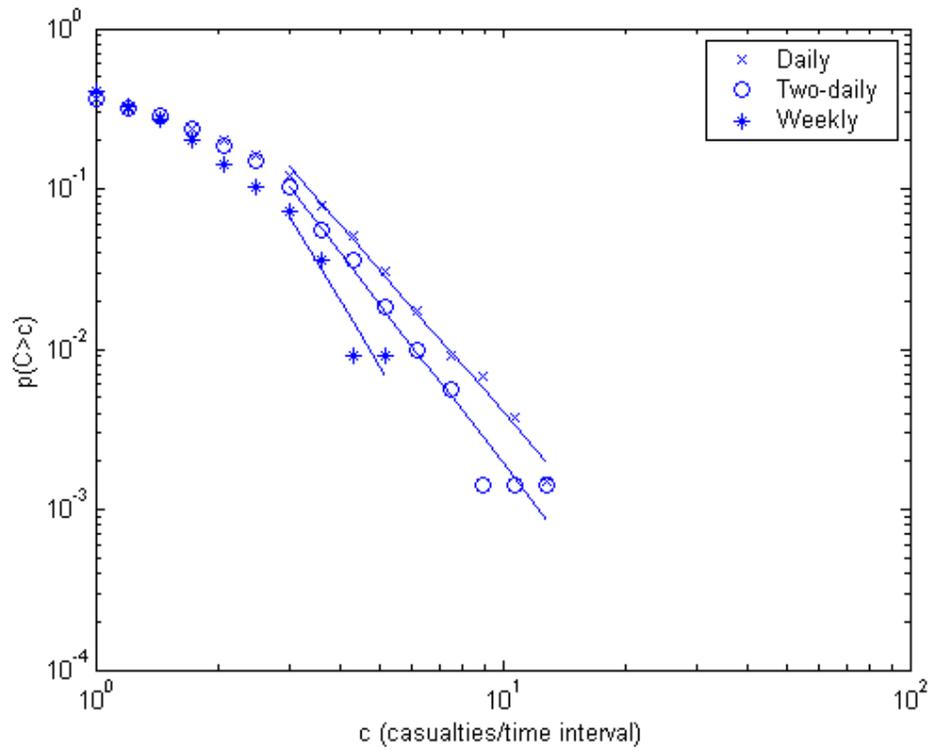

**Figure 5: Probability that an arbitrary casualty level *C* will exceed *c*, using three different levels of aggregation.**

Nevertheless, given that most operations tend to have the same sorts of things going on within them, and that any difference in the fractal dimension of the data from operation to operation may be much smaller than other uncertainties for the operation, it may be reasonable to assume that we can treat the power exponents of casualty data as a constant across sets of battles fought with similar equipment. This being so, having obtained the appropriate fractal dimension from the World War II data examined here, we need only a single parameter (the mean daily casualty rate) to estimate the likelihood of various daily casualty levels for any other similar operation.

It is beyond the scope of this report to examine particular cascade models. Instead, the method used to estimate the probability of a given casualty level is a simple approach based on the idea that multifractal data display hyperbolic tails for their exceedence probability distributions. This idea is discussed in more depth in Harris *et al.* (1996), who demonstrate the existence of hyperbolic tails for multifractal rain data. For data that behave this way, one expects that the exceedence probabilities for the data (i.e. $Pr(C_i > c)$), where $c$ is some arbitrary casualty level) will tend to a power law.

To test this for the WWII data, casualty data from several different divisions were normalized to have a mean of 1, and used as a single ensemble of data. This gave a data set of 1614 points. The exceedence distribution for this ensemble was then plotted in Figure 5. It can be seen from the figure that above a certain value for $c$, the exceedence probability distribution for the daily casualty data indeed tends to a power law, with a slope of approximately –2.9.

From the plot we can pick the point where this power law begins. This is for a $c$ value of 3.0 (i.e. 3 times the mean casualty level). This corresponds to a probability of exceedence of 0.13 (i.e. the probability of $C_i$ exceeding 3 is 13%). If we assume that the power law describing the exceedence function has a form such as:

$$P(C_i > c) = E\, c^{-2.9} \qquad (5)$$

where $E$ is a constant, then the point where the power law begins gives us the value of $E$. Substituting the probability of exceeding 3 casualties into Equation 5 gives $E = 3.1$. Then rearranging Equation 5 gives us the casualty level below which we can expect a given percentile $P$ to fall, i.e.:

$$c = \left(\frac{P}{3.1}\right)^{-0.34} \qquad (6)$$

For example, we may want to know what casualty level we could expect 95% of our daily casualty reports to fall below. Substituting $P = 0.05$ into Equation 6 gives a value of 4.1. If we know what the likely mean daily casualty rate is, we can estimate the level below which casualties fall on 95% of days by multiplying this mean by the value we obtained from Equation 6.

As an example, the 2$^{nd}$ Infantry Division suffered mean casualties of 35.6 per day during the period of the data shown here. Using our estimate, we can say that for 95% of days, we expect casualties to be below 35.6 x 4.1 = 146. Looking at the actual data, we find that for 95% of days casualties fell below 130.

Table 1 shows calculations for other percentile levels and for other divisions from Normandy. From the table, it can be seen that the 90 percentile estimates are quite close to the actual data, with the estimates becoming worse (though still in the ball park) for higher percentile estimates. Note that we should not expect the percentile estimates and the actual percentiles to match exactly, due to the limited number of data points available for the estimate. This is particularly so for the 99 percentile estimate, since this is based on just 16 points for the entire dataset, and so just two or three points for the divisional estimates.

Similar estimates can be made for the likelihood of given casualty levels for longer periods. The power exponents for Equation 6 can be obtained by finding the power-law slopes for the two-daily and weekly data in Figure 5. Since the slope becomes steeper as the resolution is decreased, the "tails" of the distributions become less extreme for the coarser-grained data. From the figure it can be seen that the value for the power exponent in Equation 6 becomes –3.6 for two-daily estimates, and –4.2 for weekly estimates.

Using a weekly casualty estimate as an example, we estimate $E = 1.7$ from Figure 5, and substitute this value and the power exponent –4.2 obtained from finding the slope into Equation 6. Taking the mean daily casualty rate for the 2$^{nd}$ Infantry division, it is expected that 90% of the time the casualties for a given week will fall below 491.

**Table 1: Values for percentile daily casualty estimates.**

| Percentile | Normalized estimate | Actual 1$^{st}$ Inf Div estimate (actual) | Actual 2$^{nd}$ Inf Div estimate (actual) | Actual 4$^{th}$ Inf Div estimate (actual) | Actual 2$^{nd}$ Arm Div estimate (actual) |
|---|---|---|---|---|---|
| 90% | 3.1 | 102 (92) | 114 (88) | 213 (186) | 64 (64) |
| 95% | 4.1 | 130 (100) | 146 (130) | 270 (253) | 81 (95) |
| 99% | 7.8 | 224 (159) | 250 (319) | 467 (470) | 139 (160) |
| Mean | 1.0 | 31.8 | 35.6 | 66.4 | 19.8 |

## 4. CONCLUSIONS

The analysis presented here demonstrated that World War II casualty data display statistical structure and probability distributions that would be expected from multifractal type data. In fact, the statistical behaviour of the casualty data was extremely similar to that observed for rainfall data by other workers.

Given that the data displayed these properties, it was shown how the existence of power-law tails in the exceedence probability distributions could be exploited to allow estimates of the likelihood of various casualty levels. This was done by normalising casualty data from several different infantry and armoured divisions, and finding the power exponent describing the tail of the exceedence probability distribution.

If it can be assumed that all such divisions fight in similar ways, and so obey similar power laws, then one only needs an estimate of the mean daily casualty rate to be able to calculate the probability of a given level of casualties for a particular period. Estimates of mean casualties can be obtained from historical analysis by other workers, such as Helmbold (1995).

However, while it appears that the assumption that the power exponent describing the probability distribution does not change (or changes only slightly) from division to division for the data analysed here, it is not clear if this is generally valid. Further investigation is needed on this point.